\documentclass[useAMS,usenatbib]{mn2e}
\usepackage{graphicx}
\usepackage{latexsym}
\usepackage{url}
\usepackage{psfig}

\title[The  semi-major axis distribution of exoplanets]{The link between disc dispersal by photoevaporation and
  the semi-major axis distribution of exoplanets}

\author[Ercolano, Rosotti]{Barbara Ercolano$^{1,2}$, Giovanni
  Rosotti$^{1,2,3}$\thanks{E-mail: ercolano@usm.lmu.de (BE)}\\
$^{1}$Universit\"ats-Sternwarte M\"unchen, Scheinerstr. 1, 81679 M\"unchen, Germany\\
$^{2}$Excellence Cluster Origin and Structure of the Universe,
Boltzmannstr.2, 85748 Garching bei M\"unchen, Germany\\
$^{3}$Institute of Astronomy, Madingley Rd, Cambridge, CB3 0HA, UK}

\begin{document}

\pagerange{\pageref{firstpage}--\pageref{lastpage}} \pubyear{2011}

\maketitle

\label{firstpage}

\def\mnras{MNRAS}
\def\apj{ApJ}
\def\aap{A\&A}
\def\apjl{ApJL}
\def\apjs{ApJS}
\def\bain{BAIN}
\def\araa{ARA\&A}
\def\pasp{PASP}
\def\aj{AJ}
\def\pasj{PASJ}
\def\ga{\sim}
\voffset-.4in

\begin{abstract}

We investigate the influence of photoevaporation of protoplanetary discs on the final distribution of exoplanets semi-major axis distances. We model giant planet migration in viscous discs affected by photoevaporation driven by either pure EUV or soft X-ray radiation (XEUV). We show that the final exoplanet distributions are strongly dependant on the choice of the photoevaporation model. In particular, we find that XEUV is more efficient than pure EUV radiation at parking planets at approximately 1-2 AU distance from their central star, hence roughly reproducing the observed peak in the exoplanets semi-major axis distributions. We note however that a more quantitative comparison with the observations is hindered by the oversimplified treatment of planetary accretion, which severely affects migration rates. For this reason, caution should be used when using these models to constrain details of disc clearing and/or migration from the observations. Nevertheless our results indicate that disc dispersal by photoevaporation may be the main driver of the features in the exoplanets semi-major axis distribution observed by recent surveys.  

\end{abstract}


\section{Introduction}

Statistical studies of the properties of extrasolar planets have recently become possible thanks to the discovery of over 1500 confirmed planets and almost 5000 total planets, if the Kepler candidates are included (exoplanets.org, Han et al. 2014). In particular the study of the distribution of exoplanets semi-major axis could provide useful insights into the planetary migration and protoplanetary disc evolution process (e.g. Alexander \& Pascucci, 2012, from now on AP12). The occurrence of planetary migration is reasonably well accepted for high mass planets, which migrate inwards due to tidal interactions with the protoplanetary disc (e.g. Goldreich \& Tremaine 1980; Lin \& Papaloizou 1986). Indeed a number of numerical models have been computed, which roughly agree with the observed distributions (e.g. Armitage 2007; Alexander \& Armitage 2009; AP12). The migration of planets larger than 0,5 M$_{\rm Jup}$ follows the so-called Type II migration, which sets in when a gap in the protoplanetary disc is dynamically cleared by the planet itself. At this point the planet and the gap keep moving inwards due to the material which keeps flowing into the gap on timescales dictated by viscous accretion of the protoplanetary disc. While gas is present in the disc, this process is unstoppable and irreversible, meaning that without a disc clearing mechanism all giant planets would be expected to fall into the star at some point. Indeed, planetary population synthesis models (Benz et al. 2014) include simplified prescriptions for eventually dissipating the disc. Photoevaporation of the gas disc by high energy radiation from the central star (e.g. Alexander et al. 2006; Gorti et al. 2009; Ercolano et al. 2009; Owen et al. 2010) provides a mechanism to remove the gas from the disc and cease migration.

AP12 construct numerical models of planet migration in evolving protoplanetary discs in order to investigate the effect of disc clearing on the final semi-major axis distribution of exoplanets. They find that assuming a pure EUV-photoevaporation profile (Alexander et al. 2006) and a flat mass distribution of extra-solar planet the observed peak at semi-major axis distances of approximately 1-2 AU could be roughly reproduced (see Hagesawa \& Pudritz for an alternative explanation for the origin of planetary system architectures). However, depending on the nature of the photoevaporating radiation (EUV, X-ray or FUV) the shape of the photoevaporation profile, as well as the photoevaporation rate itself, change dramatically. For example, combined X-ray and EUV (XEUV) photoevaporation rates are roughly 100 times larger than pure EUV photoevaporation rates (Owen et al. 2010), and  XEUV photoevaporation rates as a function of disc radius are not so strongly concentrated at the gravitational radius as in the pure EUV case (see e.g. Armitage 2011). One could expect that these large differences should result also in different final distributions of exoplanets semi-major axes, meaning that observed distributions may provide a constraint on the dominant disc dispersal radiation. Furthermore the initial planet mass function may also have an influence on the final exoplanet semi-major axes distribution. 

In this paper we perform numerical modelling to explore the influence of different photoevaporation models and of the initial planet mass function on the final distributions of exoplanets semi-major axis, in order to explore the possibility to use the latter to constrain the initial conditions and the nature of the photoevaporating radiation. In Section 2 we describe our numerical approach and present a number of tests of the new code we developed to this aim. In Section 3 we present and discuss our results, while a summary of our findings is given in Section 4. 


\section{Modelling Approach}

Our modelling approach is similar to that described in AP12, based on
the numerical model of Alexander \& Armitage (2009). We summarise here
the basic concepts for convenience. Each model includes a
protoplanetary disc which evolves under the influence of viscosity in
the disc and photoevaporation from its central star. After a given timescale $t_\mathrm{form}$, a giant planet of
a given mass $m_\mathrm{p}$ is assumed to form within the disc and
undergoes Type II migration due to the torques from the gaseous disc. Angular momentum conservation dictates that the disc evolution is in turn affected by the planet. We do not model the formation of the planet, and assume that this is formed istantaneously in the disc. The disc-planet system evolves
according to the following equation:

\begin{equation} \label{eq:sigma_evol}
\frac{\partial \Sigma}{\partial t} = \frac{1}{R}\frac{\partial}{\partial R}\left[ 3R^{1/2} \frac{\partial}{\partial R}\left(\nu \Sigma R^{1/2}\right) - \frac{2 \Lambda \Sigma R^{3/2}}{(GM_*)^{1/2}}\right] - \dot{\Sigma}_{\mathrm {w}}(R,t) \, .
\end{equation}
where the first term describes the viscous evolution of the disc (Lynden-Bell \& Pringle 1974), the second term the migration of the planet due to the torques from the disc (e.g. Lin \& Papaloizou 1986) and the last term the mass loss due to photoevaporation (e.g. Clarke et al. 2001). Here $\Sigma$ is the surface density of the disc, $R$ the distance from the star in the plane of the disc, $\nu$ the kinematical viscosity of the disc, $M_\ast$ the mass of the star and $\dot{\Sigma}_{\mathrm {w}}$ the photoevaporation profile. 

The term $\Lambda$ is the rate of specific angular momentum transfer from the planet to the disc. We use the following form for $\Lambda$, which is a slight modification by Armitage et al. (2002) of the original one used by Lin \& Papaloizou (1986):
\begin{equation}
\Lambda(R,a) = \left\{ \begin{array}{ll}
- \frac{q^2 GM_*}{2R} \left(\frac{R}{\Delta_{\mathrm p}}\right)^4 & \textrm{if } \, R < a\\
\frac{q^2 GM_*}{2R} \left(\frac{a}{\Delta_{\mathrm p}}\right)^4 & \textrm{if } \,R > a\\
\end{array}\right. ,
\end{equation}
where $q$ is the mass ratio between the planet and the star, $a$ is the semi-major axis of the planet orbit (assumed to be circular) and $\Delta_p$ is given by
\begin{equation}
 \Delta_\mathrm{p} = \max (H, |R-a| ),
\end{equation}
where $H$ is the disc scale height.

Due to the disc torques, the planet migrates at a rate given by:
\begin{equation}
 \frac{da}{dt}= - \left( \frac{a}{GM_\ast} \right)^{1/2} \left( \frac{4\pi}{M_p} \right) \int_{r_\mathrm{in}}^{r_\mathrm{out}} r \Lambda \Sigma dr
\end{equation}

This approach for describing migration is computationally more efficient than more complicated treatments, yet it was shown to give comparable results (Takeuchi et al., 1996, Trilling 1998). In Type II migration, the planet behaves essentially like a test particle entrained in the viscous flow, and this is the fundamental reason why a more detailed description does not yield significantly different results. Note, however, that doubts have been cast  recently on the correctness of Type II migration in hydrodynamical simulations (Duffell et al. 2014, D{\"u}ermann \& Kley 2015). We ignored these recent results as they are still somewhat uncertain, but note that they might point to important effects to be included in future works.

To solve equation \ref{eq:sigma_evol}, we use standard numerical techniques. We discretise the equation on a grid of points equispaced in $R^{1/2}$. The viscous term is treated as a diffusive term (Pringle \& Verbunt 1986). The planet torque is treated as an advection term, and computed after the diffusive term using operator splitting. We use the van Leer (1977) method to reconstruct the surface density at the cell boundaries. The photoevaporation term is integrated by simply removing a fixed amount of mass from each cell at every timestep. To prevent numerical problems, we use a floor surface density of $10^{-8} g/cm^2$. Following Armitage et al. (2002) we limit the maximum planet torque close to the planet for computational reasons; this has no consequences on the orbital migration rate. We use a resolution of 4000 points and our grid extends from 0.04 to $10^4 \ \mathrm{AU}$. This is higher than what commonly employes in similar treatments. 

\subsection{Planetary accretion}

It is well known that gaps cleared by giant planets are not able to stop completely the mass flow from the outer disc (e.g., Lubow \& D'angelo 2006). For this reason we follow partially Alexander \& Armitage (2009) and include a prescription, calibrated on published results from hydrodynamical simulations, to include this mass flow. We use the same prescription from AP12, which was proposed by Veras \& Armitage (2004):
\begin{equation}
 \frac{\epsilon(M_p)}{\epsilon_\mathrm{max}} = 1.67 \left( \frac{M_p}{1 M_\mathrm{jup} } \right)^{1/3} \exp \left( - \frac{M_p}{1.5 M_\mathrm{jup}} \right) + 0.04.
\end{equation}
Here, the efficiency $\epsilon (M_p)$ is defined as the ratio between the accretion rate onto the planet and the accretion rate in a steady disc in absence of the planet. It follows that
\begin{equation}
 \dot{M}_p = \epsilon(M_p) \dot{M}_{disc}.
\end{equation}
Following AP12, the mass accretion rate through the gap is computed via
\begin{equation}
 \dot{M}_{inner} = \frac{1}{1+\epsilon} {\dot{M}_p}.
\end{equation}

Note, however, that the sum of the two rates is not $\dot{M}_{disc}$, and therefore the correct rate to apply at the outer edge of the gap to enforce mass conservation is the sum of the two rates. We note that such parameterisation may be avoided by using the results of Lubow \& D'angelo (2006), who showed via analytical arguments that $\dot{M}_{inner} = \dot{M}_p /\epsilon$. We chose however to follow AP12 for the sake of comparison, and note that given the current uncertainties in these rates, both prescriptions may carry large uncertainties. 

To apply this mass leakage through the gap computationally, we set the cells outside the orbit of the planet to the floor value, until we have removed the mass that is flowing in the timestep. We then use the prescribed rate $\dot{M}_p$ to increase the mass of the planet and we ``unload'' the mass coming from the rate $\dot{M}_{inner}$ in the first cell inside the orbit of the planet. This algorithm follows from a private communication with R. Alexander and was chosen for consistency with their results. Possible improvements would consist in removing and adding the mass from a larger region, to smooth the effect of this procedure (e.g., Owen 2014).

\subsection{photoevaporation modelling}
Several photoevaporation models are present in the literature, considering different irradiating fields. The EUV field was the first to be explored (Hollenbach et al. 1994) through analytical calculations. The original estimates were then refined through hydrodynamical simulations by Font et al. (2004). Finally, Alexander et al. 2006 have shown the importance of including the direct illumination from the central star once a hole has been opened in the disc. We follow AP12 and assume a constant EUV flux $\Phi = 10^{42}$ photons per second in our models. We note however that this parameter is currently poorly constrained, and recent observational results (Pascucci et al. 2014) imply that this may have been significantly overestimated. The total mass-loss rate with this flux is $4 \times 10^{-10} \ M_\odot \ \mathrm{yr}^{-1}$ (Alexander et al. 2006). The mass-loss profile $\dot{\Sigma}(r)$ can be found in the appendix of Alexander \& Armitage (2007). We distinguish between a \textit{diffuse} profile, which is the relevant one for most of the time while there is a full disc, and a \textit{direct} profile, which is the relevant one after a hole has opened in the disc and the inner edge of the outer disc is directly exposed to the EUV radiation. A criterion is needed to switch between the two. AP12 apply the switch when the surface density in the inner disc falls below a critical value $\Sigma_c = 10^{-5} \ \mathrm{g} \ \mathrm{cm}^{-2}$. Following a private communication with R. Alexander, we monitor the surface density at two different locations in the disc, at $0.1 \ \mathrm{AU}$ and $1 \ \mathrm{AU}$ from the star, and switch between the two profiles when both have fallen below the critical value. The exact criterion assumed is important, since when the direct profile is applied the disc is dispersed in a short time-scale, and the planet migration is essentially stalled.

In addition to the original EUV model used by AP12, we employ here also the XEUV model of Owen et al. (2010, 2011, 2012). In this case the mass-loss rate is given by the following relation (Owen et al. 2012):
\begin{equation}
\dot{M} = 6.25 \times 10^{-9} \ M_\odot \ \mathrm{yr}^{-1} \ \left(\frac{L_\mathrm{X}}{10^{30} \ \mathrm{erg} \ \mathrm{s}^{-1}} \right)^{1.14}
\end{equation}
where $L_\mathrm{X}$ is the X-ray luminosity of the star. The X-ray luminosity function is well characterized by observations (e.g. Preibisch et al. 2005 and Guedel et al. 2007), and the median is $1.1 \times 10^{30} \ \mathrm{erg} \ \mathrm{s}^{-1}$ for solar type stars. Note that the <mass-loss rate corresponding to this X-ray luminosity is almost 100 times higher than the EUV one. The shape of the $\dot{\Sigma}(r)$ function is parameterised in appendix A of Owen et al. (2012).

Finally, a photoevaporation model based on FUV radiation has been presented by Gorti, Hollenbach \& Dullemond (2009). However, this work does not include a hydrodynamical solution for the wind, meaning that the resulting rates are at this time still uncertain and will not be considered here. 


\subsection{Population synthesis}

We construct a population synthesis of disc-planet systems by randomly
sampling a number of initial conditions, which are summarised for the two models in Table \ref{tab:disc_prop}. The two different photoevaporation models that we consider have very different mass-loss rates. This would lead to differences in the lifetimes of the disc if the same initial conditions are assumed. To reconcile the population synthesis with the observed disc lifetimes, the two different photoevaporation models need thus to assume different initial disc properties. The EUV model, which has a lower photoevaporation rate, assumes a slighly less massive, faster evolving disc than the XEUV one. Note that, since both the disc viscosity and the disc masses are poorly constrained, both models are a reasonable description within the current uncertainties. In all of our models we assume the initial surface density profile to be
\begin{equation}
 \Sigma (R,t=0) = \Sigma_0 (\frac{R_0}  {R}) \exp(-\frac{R}{R_0}),
\end{equation}
where $\Sigma_0$ is a normalization constant that we fix by imposing a given total mass of the disc, and $R_0$ is the initial size of the disc. We assume that the viscosity profile follows  $\nu \propto R$. This is convenient from the theoretical point of view, as it corresponds to a mildly flared disc and a constant value of $\alpha$ throughout the disc. In what follows we will speak of the normalization of the viscosity profile through the viscous time $t_\nu (R) = R^2/3\nu(R) \propto R$.

In the EUV model, we follow AP12 in varying the initial disc mass. We assume a log-normal distribution, with a median of $10^{-1.5} \ M_\odot$ and a $3 \sigma$ dispersion of $0.5 \ \mathrm{dex}$. The viscous time is $5 \times 10^4 \ \mathrm{yr}$ at $10 \ \mathrm{AU}$. Note that in 1d models there is a degeneracy between $\alpha$, the Shakura-Sunyaev parameter, and $H/R$, the aspect ratio of the disc. The viscous time is the relevant quantiy that controls the evolution of the disc, and not the $\alpha$ parameter. For reference, we note that assuming a value of $\alpha=0.01$ and an aspect ratio of $H/R=0.0333$ at $1 \ \mathrm{AU}$ gives the assumed viscous time. We follow AP12 in assigning the properties of the formed giant planets. Rather than modelling the formation process, we prescribe statistically the properties of the formed planets, as the purpose of this exercise is to assess the implications of disc dispersal, rather than building a fully consistent planetary population synthesis model. We assume that all planets form at $5 \ \mathrm{AU}$, with a mass assigned from a uniform distribution between $0.5$ and $5 \ M_\mathrm{jup}$. We will later correct statistically this assumption. The other important variable is the formation time, which we also draw randomly from an uniform distribution between $0.25 \ \mathrm{Myr}$ and $t_c=1/3 t_\nu (3M_d/ (2t_\nu \dot{M}_w) )^{2/3}$ (Clarke et al. 2001, Ruden 2004), which is the time of disc clearing.

In the XEUV model, we follow Owen et al. (2011) in assuming that all discs start with the same initial conditions, and the only difference in their evolution is due to the spread in X-ray luminosities. Here we assume an initial mass of $0.07 \ M_\odot$, and a viscous time of $7 \times 10^5 \ \mathrm{yr}$ at $18 \ \mathrm{AU}$. We use the X-ray luminosity function from Guedel et al. (2007), obtained for the Taurus cluster, to prescribe the X-ray luminosity of the star, which sets the mass-loss rate. The properties of the formed planets are assigned as before.

\begin{table*}
\begin{tabular}{c|c|c}
\hline
Model & EUV & XEUV \\
\hline
Disc mass & median $10^{-1.5}$M$_{\odot}$, log-normal with variance 0.5 dex  & 0.07 M$_{\odot}$ \\
Scaling radius & 10 AU & 18 AU \\
Star mass & 1 $M_\odot$ & 0.7 $M_\odot$ \\
Viscous time at 10 AU & $10^4$ yr & $4 \times 10^{5}$ yr\\
X-ray luminosity & / & X-ray LF from Taurus, Guedel et al. (2007) \\
EUV flux & $10^{42}$ phot/yr & / \\
\hline
\end{tabular}
\caption{Disc properties in the two models}
\label{tab:disc_prop}
\end{table*}


\subsection{Code tests}

\begin{figure}
 \includegraphics[width=0.47\textwidth]{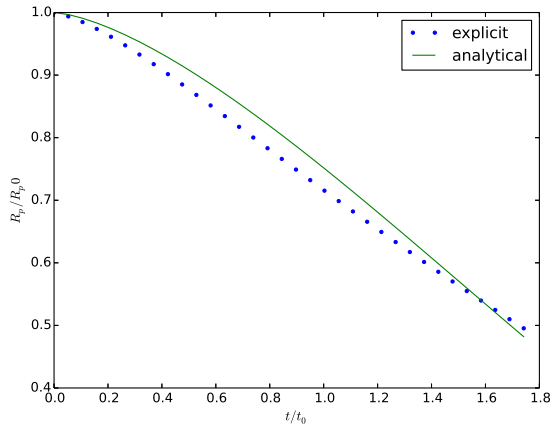}
\caption{Comparison Spock against Ivanov analytical solution}
\label{fig:comp_ivanov}
\end{figure}

\begin{figure}
\begin{center}
\includegraphics[width=0.47\textwidth]{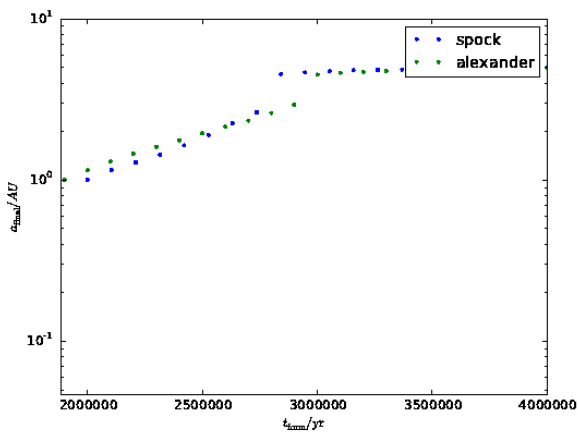}
\caption{Comparison Spock against AP12}
\label{fig:comp}
\end{center}
\end{figure}

We have tested independently the different components of our code, Spock. The viscous evolution part is the easiest to check, since analytical solutions (Lynden-Bell \& Pringle 1974) are readily available. No significant deviation has been observed. We tested the implementation of the photoevaporation profiles together with the viscous evolution by checking that the lifetimes of the discs (without planets) reported by Alexander et al. (2006,2009) and Owen et al. (2010) are correctly recovered.

The planet migration part can be tested against an existing analytical solution (Ivanov 1999). A comparison is shown in Figure \ref{fig:comp_ivanov}. It should be noted that the analytical solution employs a different treatment for the planet, and therefore an exact overlap between the two is not to be expected. Our result is similar to other codes (e.g., see Figure 3 in Lodato \& Clarke 2004 for a similar test).

Figure~\ref{fig:comp} shows a comparison run of our code with
AP12 for a disc of mass $10^{-1.5} M_\odot$. The two codes show reasonable agreement, particularly when
considering that a number of relatively unconstrained parameters have
to be introduced in the model, which may influence the final
result. We have however no way of telling if those differences may
become more serious for planets of different masses, as AP12 runs at
different masses are not available. We fear that this may be indeed
the case when we compare the final distributions of exoplanets we
obtain from our code for the same initial conditions parameter range
as AP12, as shown by the red dotted lines in the left and right panels of 
Figure~\ref{fig:xeuvhighlowmass}.Both for the high (right) and low (left) mass exoplanet distributions, we do not find 
the strong peaks reported by AP12 at semimajor axis values between 1 and 2~AU,
indeed our distributions for the EUV case are relatively flat. 
Without a dedicated benchmarking effort of the two codes, the exact
cause of this discrepancy is unknown, but it probably lies in the 
(often arbitrary) choices of a number of parameters describing
the redistribution of mass due to accretion on the planet and the leakage through the gap, which in
turn has influence on the migration rate of the planet itself. The other very sensitive part of the code is the switch between the direct and diffuse profile, which rapidly dissipates the disc and is therefore very important to treat correctly. Until progresses are made in the physical description of planetary accretion and migration, the predicting power of models like those presented here or in AP12 is severely 
reduced and hence their use as diagnostics of disc clearing, planet migration
and planetary accretion should be treated with caution. 

Nevertheless the main aim of this work is not affected by these uncertainties, as we wish to investigate the influence of the photoevaporation model and the exoplanet mass function on the final exoplanet distribution. These aims can be achieved regardless of the uncertainties inherent to the modelling that are highlighted above, as long as selfconsistence within the same code is observed. 
\section{Results}


Following AP12, in our further discussion we divide our planetary sample into two mass bins: the low mass planets comprising planets of masses from 0.5~M$_{\odot}$ to 2.5~$M_{\odot}$ and a high mass bin comprising planets with M$_{\odot} >  2.5 M_{\odot}$. We explore in what follows the importance of the chosen photoevaporation model and initial planet mass function assumed on the final semi-major axis distribution of exoplanets. 

\subsection{Effects of photoevaporation models on the final semi-major axis distribution of exoplanets}

\begin{figure*}
\begin{center}
\includegraphics[width=0.47\textwidth]{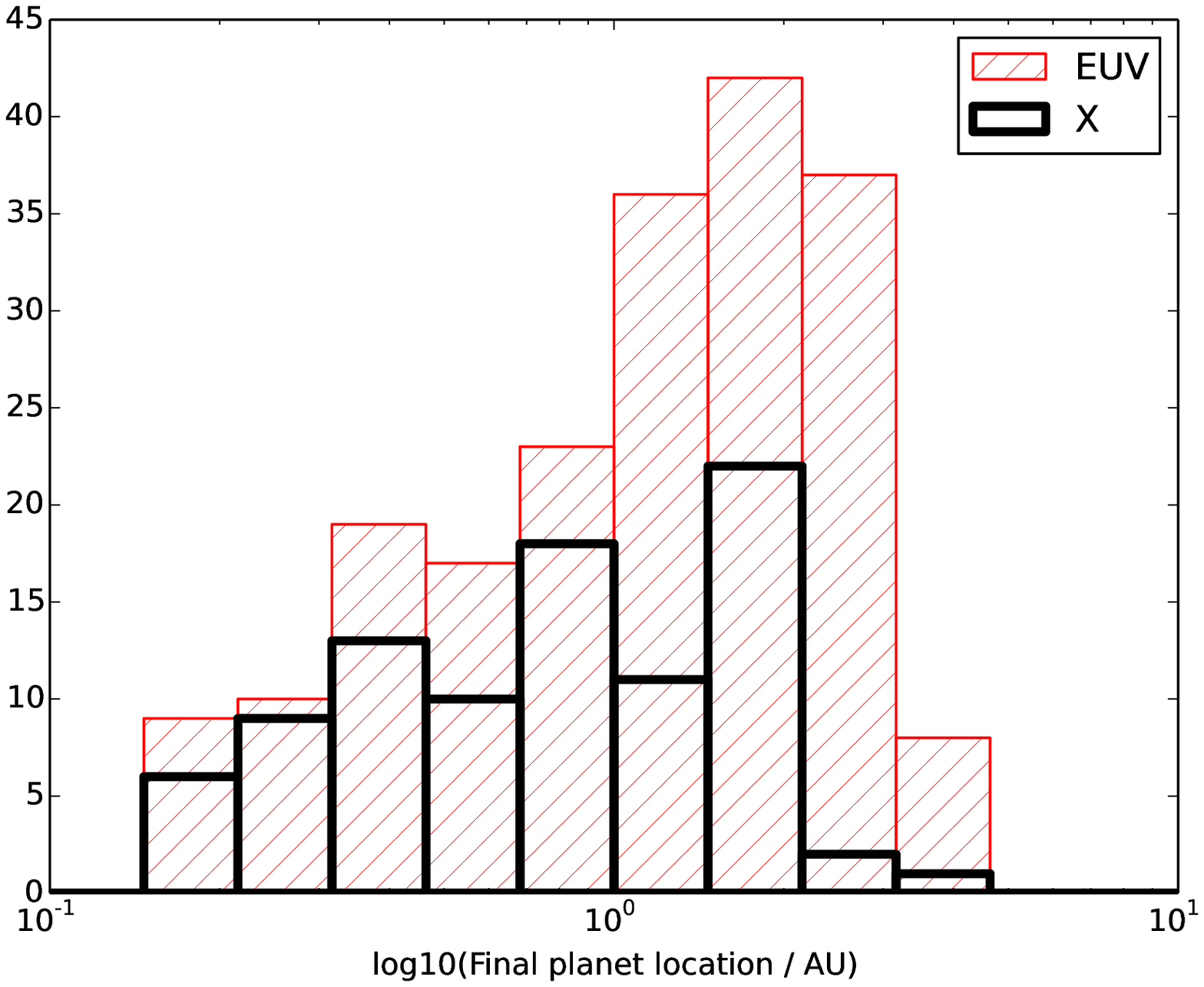}
\includegraphics[width=0.47\textwidth]{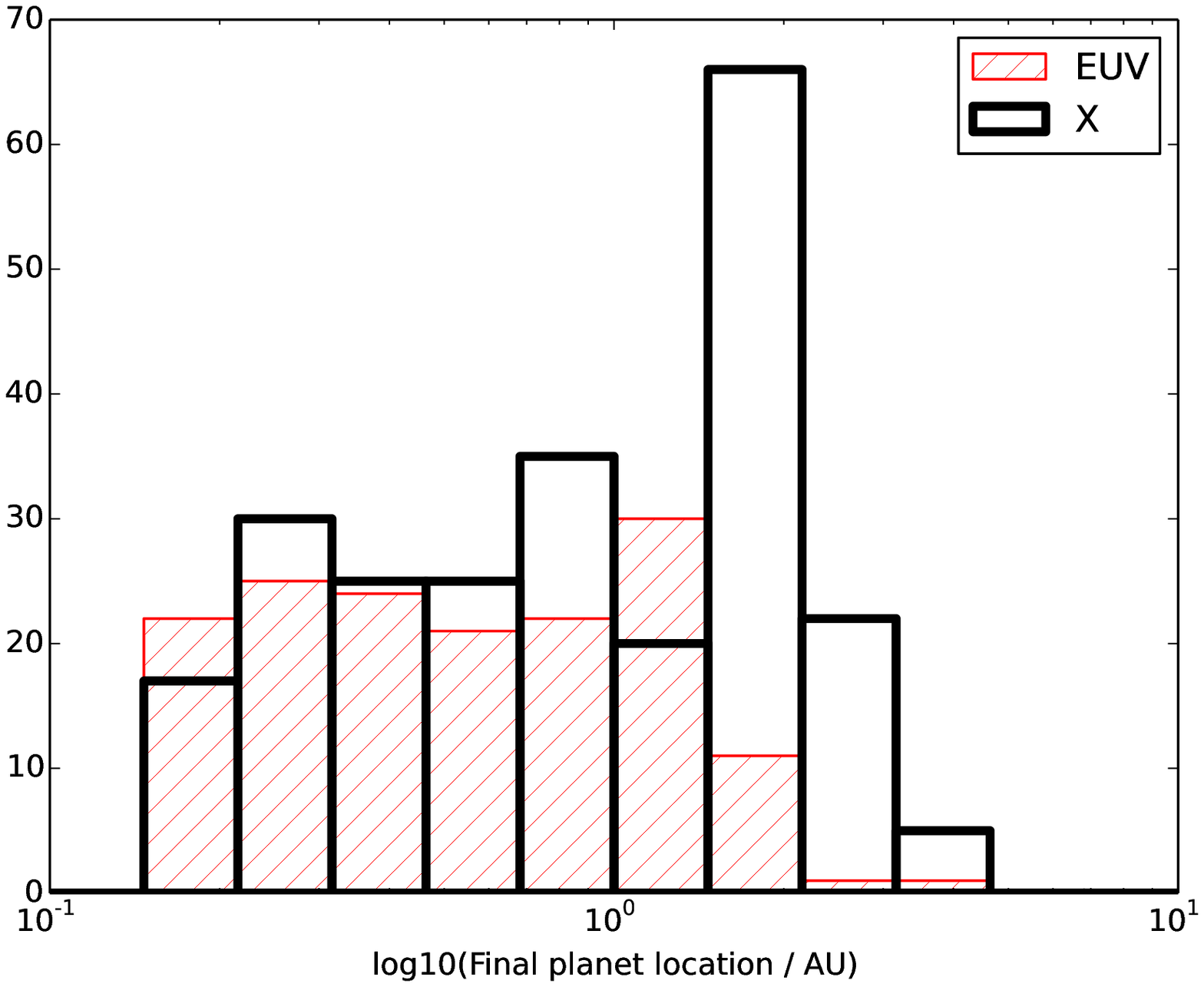}
\caption{Histogram of exoplanet semi-major axes distribution for low -- 0.5~M$_{\odot} < M_p <  2.5~M_{\odot}$-  (left panel) and high mass stars -- $M_p >  2.5 M_{\odot}$-  (right panel). The black and red dotted lines are for the XEUV and EUV cases, respectively. A flat initial mass distribution of planets is assumed. }
\label{fig:xeuvhighlowmass}
\end{center}
\end{figure*}

The left and right panels of Figure~\ref{fig:xeuvhighlowmass} show a comparison of the final semi-major axis distribution of the low and high mass  exoplanets formed in discs under the influence of either EUV or XEUV photoevaporation. The distributions shown in the Figure were obtained assuming a flat initial mass function of exoplanets.  As mentioned already in the previous section, the distributions obtained for discs under the influence of pure EUV photoevaporation are rather flat. The XEUV case shows on the contrary peaked distributions, with the peak being at semi-major axes values roughly comparable to those observations. It is worth keeping in mind, when comparing the XEUV to the EUV case that the accretion properties of the discs in these two cases are very different. In order to obtain similar disc lifetimes, the XEUV case requires a larger accretion rate to disc mass ratio than the EUV case, this leads to faster accretion of mass onto the planets in the XEUV case, and the overpopulation (about 800 planets) of the high mass bin (right panel in Figure~\ref{fig:xeuvhighlowmass}) compared to the EUV case (about 150 planets). Once a planet reaches a given mass, the flow of material through the gap is strongly reduced, aiding the onset of a PIPE-like dispersal (Rosotti et al. 2013), i.e. inner disc dispersal by photoevaporation, which stops further migration of these planets. The final result is that a much larger number of high mass planets are retained in the XEUV case, compared to the EUV. Indeed both models start with 250 planets in this mass bin with the EUV case ending with only 150 planets in this mass bin, compared to the 800 planets in the XEUV case. The peak seen in the XEUV case is indeed due to this parking effect brought by the fast inner disc removal. 

It is clear from the above that the final distribution of extrasolar planets semi-major axis is seriously affected by the choice of photoevaporation model. As discussed above the main driver of the strong differences, however, is the different total mass loss rates predicted by the two models, which lead to differen accretion properties that have to be assumed in the two cases in order to preserve the correct disc lifetimes. The question then remains as to whether the photoevaporation {\it profile} itself has an influence on the final distribution of exoplanets semi-major axes. To answer this question we have constructed a further population synthesis where we keep all mass accretion rate parameters the same as in the EUV case, but apply the XEUV mass loss profile, normalised to a total mass loss rate of $4 \times 10^{-10} \ M_\odot \ \mathrm{yr}^{-1}$, which corresponds to the pure EUV model. This ensures that the correct disc lifetimes are preserved while isolating the effect of the mass-loss profile on the final distributions. 

\begin{figure*}
\begin{center}
\includegraphics[width=0.47\textwidth]{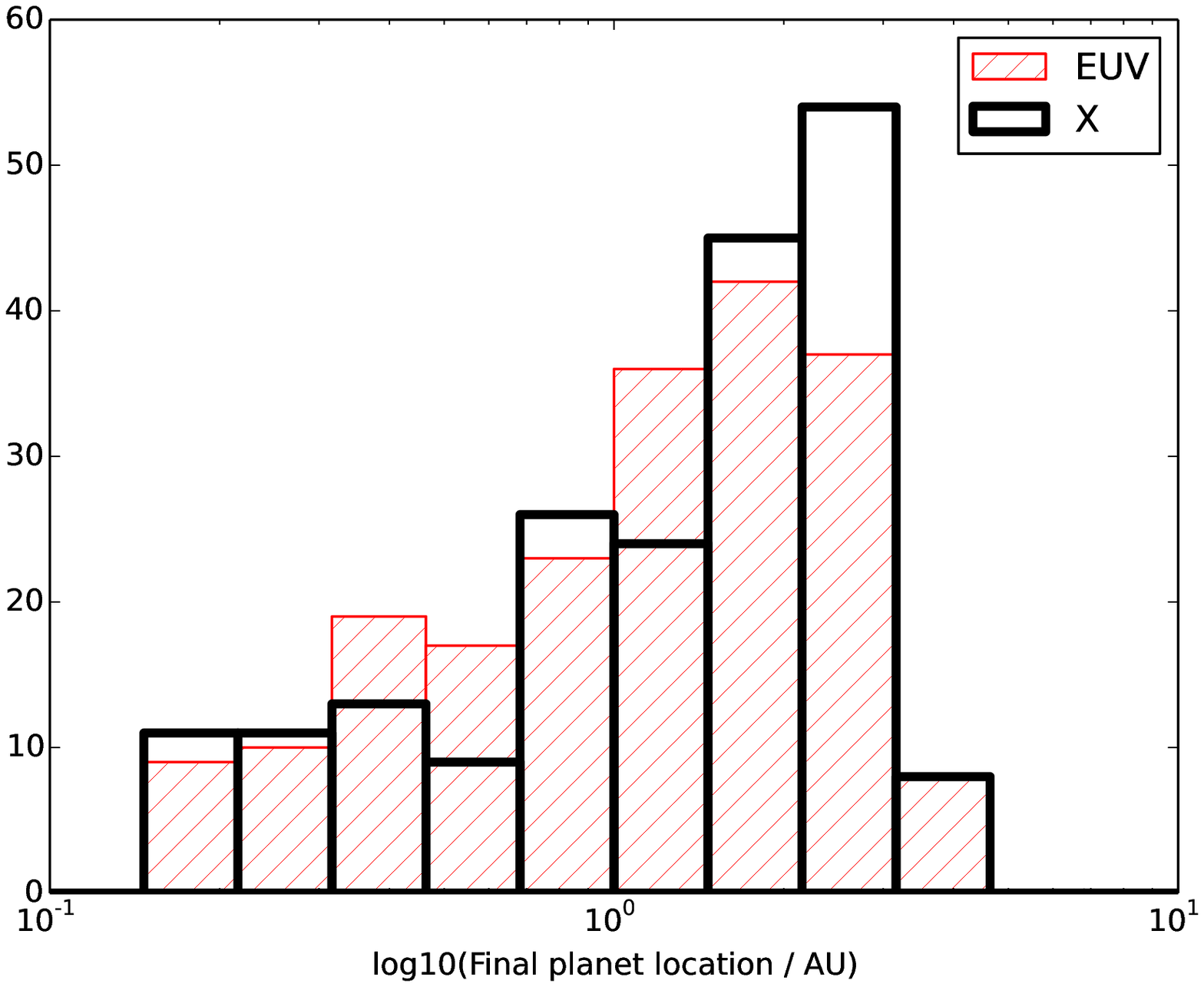}
\includegraphics[width=0.47\textwidth]{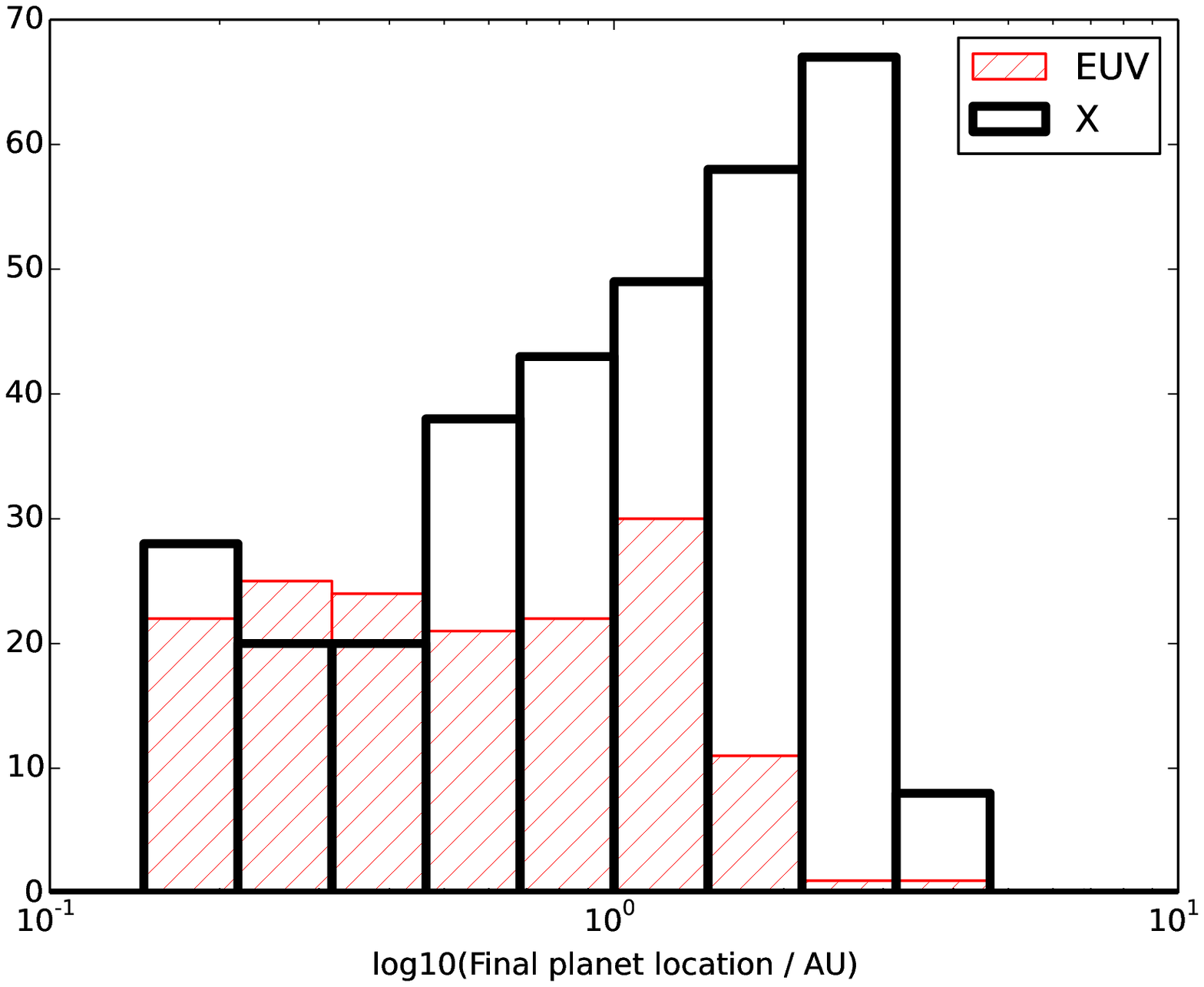}
\caption{Histogram of exoplanet semi-major axes distribution for low -- 0.5~M$_{\odot} < M_p <  2.5~M_{\odot}$-  (left panel) and high mass stars -- $M_p >  2.5 M_{\odot}$-  (right panel). The black and red dotted lines are for the XEUV and EUV profile cases, respectively. In this case only the photoevaporation profile changes (see text for details). A flat initial mass distribution of planets is assumed.}
\label{fig:profiletest}
\end{center}
\end{figure*}

The results of this academic experiment are shown in Figure~\ref{fig:profiletest}. For low mass planets very little difference is seen between the EUV and XEUV profile cases, however the situation is dramatically different in the high mass bin. Here many more planets are parked at large radii when an XEUV profile is used instead of an EUV for the same disc conditions. The disc lifetimes are very similar in the two cases, since the total mass loss rates have been set to the same (EUV) value, in fact the onset of dispersal by photoevaporation is even delayed in the XEUV case, given that for the same mass loss rate the profile is broader. The broader profile for the XEUV case leads also to migration rates for high mass planet in the XEUV profile case that are slower than in the EUV case. This happens because the XEUV is efficient at removing mass from the disc out to several tens of AU, while EUV-driven photoevaporation is mostly concentrated at the gravitational radius (see e.g. Armitage, 2011).  Therefore a planet that migrates inwards in an XEUV photoevaporated disc will be subjected to weaker gas torques as the surface densities in the disc will be lower over a broader range of disc radii. So by the time the inner disc is removed by photoevaporation, planets in an XEUV photoevaporated disc will have migrated less than planets in an EUV photoevaporated disc. Indeed those in an EUV photoevaporated disc are free to fall inwards undisturbed at least until they reach the gravitational radius of the EUV irradiated disc ($\sim$1AU which is indeed where we see a small pile up in the EUV case). It is also worth keeping in mind at this point that the migration rate through the disc is only slowed down if  $\Sigma*a_p^2 < M_p$, otherwise migration proceeds at the viscous time irrespective of the disc mass. This explains why we only observe this effect in the high mass bin.

\begin{figure*}
\begin{center}
\includegraphics[width=0.47\textwidth]{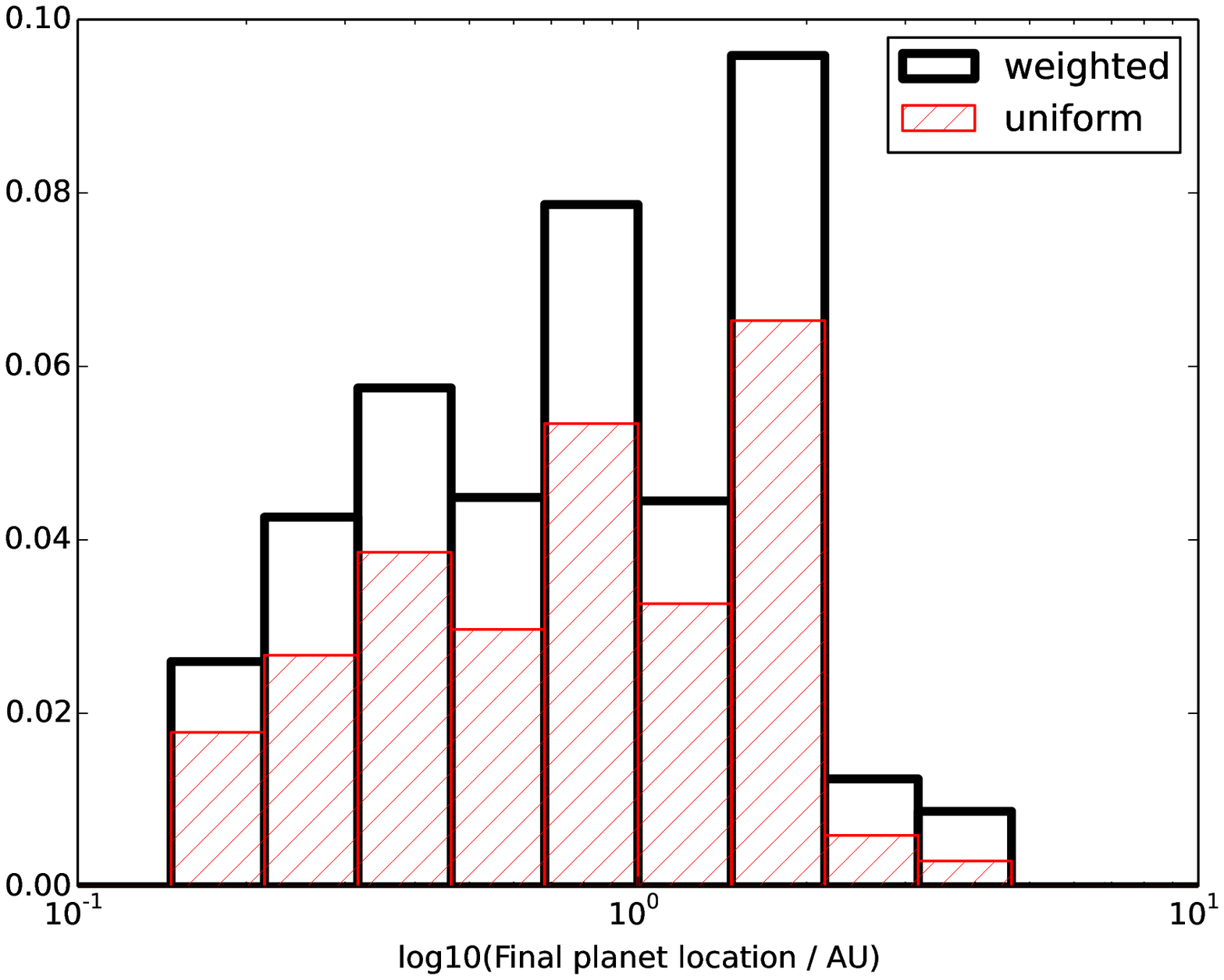}
\includegraphics[width=0.47\textwidth]{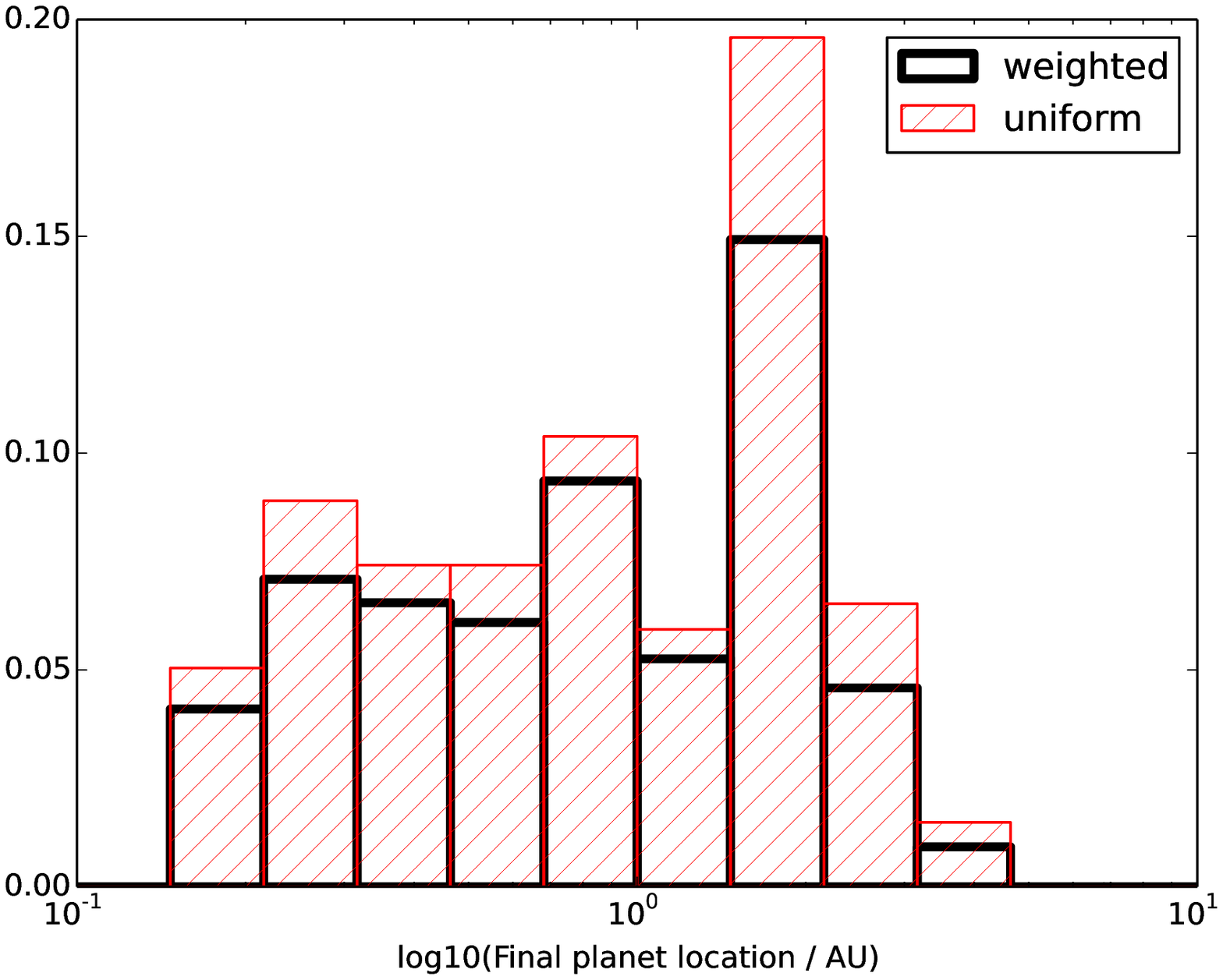}
\caption{Histogram of exoplanet semi-major axes distribution for low -- 0.5~M$_{\odot} < M_p <  2.5~M_{\odot}$-  (left panel) and high mass stars -- $M_p >  2.5 M_{\odot}$-  (right panel). The black and red dotted lines are for the XEUV case comparing, respectively, a 1/M initial mass distribution of planets  with a flat initial mass distribution.}
\label{fig:masstest}
\end{center}
\end{figure*}


\subsection{Effects of the initial mass function on the final semi-major axis distribution of exoplanets}

The exact shape of the initial exoplanet mass function is still somewhat uncertain. In the above tests and in AP12 a flat mass distribution of planets was assumed. Given the different migration rates and the different influence of higher mass planets on the transport of the material across the gap, which determines the support of the inner disc, one could suspect that the shape of the initial mass function of exoplanets should also influence the final exoplanets semi-major axis distribution. We have compared the final distributions obtained for the XEUV and the EUV cases assuming a flat initial planet mass distribution and a simple 1/M distribution and find the final distributions to be qualitatively similar in the two cases. This is shown in Figure~5 for the XEUV case. The pure EUV case yielded similar results for the two initial mass distributions and it is not shown here. Both for the EUV and XEUV case the assumption of a 1/M initial planet mass distribution yield to a slight overpopulation of planets parked at slightly larger radii. The plots were constructed considering the mass of the planets at the end of the simulation, comparing the mass of the planets injected at the beginning of the simulation increases slightly the importance of the bins at larger semi-major axis distances, but the results remain qualitatively the same. 

\section{Conclusions}

We have used a numerical approach to investigate the influence of internal photoevaporation of protoplanetary discs on the final exoplanet semi-major axis distance. We found the final distributions of exoplanets semi-major axes to be extremely sensitive to the choice of photoevaporation model in the gaseous disc. This implies that exoplanets semimajor axis measurements, which are becoming statistically significant with the more recent surveys, may be theoretically used to constrain the details of disc dispersal. Our results show that an XEUV photoevaporation profile naturally reproduces the observed peak of planets residing at roughly 1-2 AU distance from the central star. The XEUV photoevaporation model is more efficient at parking planets at larger distances from the central star than a pure EUV photoevaporation models. This happens thanks to the ability of the XEUV model at removing gas from an extended region in the disc in contrast to the pure EUV model, in which the mass loss rate is concentrated at the gravitational radius. An XEUV photoevaporated disc experiences a lowering of the gas surface density at a range of disc radii, which slows down migration particularly for the larger planets. We caution, however, that while our study aims mainly at theoretically comparing the influence of difference photoevaporation models on the final semi-major axis distribution of exoplanets, a more detailed comparison with the observations remains difficult, due to   a number of poorly constrained parameters on which the theoretical models rely. In particular, migration rates, which are crucial to the problem, are severely affected by details of the planet accretion formalism employed. In any case, our results also highlight the need of employing more precise treatments of photoevaporation in planet population synthesis models.

\section{Acknowledgements}
We thank Richard Alexander for providing details of the numerical procedures and comments on the paper. We also thank James Owen, Tilman Birnstiel and Marco Tazzari for their help and the many useful discussions. This research has made use of the Exoplanet Orbit Database and the Exoplanet Data Explorer at exoplanets.org.

\label{lastpage}

\end{document}